\documentclass[letterpaper,12pt]{article}
\usepackage{graphicx}
\usepackage[round]{natbib}
\usepackage[backref=false,hyperfootnotes=false]{hyperref}
\usepackage{multicol}

\addtocounter{secnumdepth}{1}
\addtocounter{tocdepth}{1}
\begin{document}

\begin{titlepage}
\vspace{1.5cm}
\begin{center}
\LARGE{Temporal Variability of Stars and Stellar Systems} \\
\vspace{1.0cm}
\large{A White Paper for the Stars and Stellar Evolution Science
Frontier Panel of the Astro2010 Decadal Survey} \\
\vspace{2.5cm}
{\large
Primary Author: \textbf{Tim Lister} \\
Address: Las Cumbres Observatory Global Telescope Network Inc., \\6740 Cortona 
Drive Suite 102, Santa Barbara, CA 93117, USA \\
Phone number: (805) 880 1632 \\
Email address: tlister@lcogt.net \\
\vspace{1cm}
Co authors:\textbf{ Travis Metcalfe} \\
Address: High Altitude Observatory, National Center for Atmospheric Research,
P.O. Box 3000, Boulder, Colorado 80307-3000 \\
\vspace{0.5cm}
\textbf{Tim Brown, Rachel Street} \\
Address: Las Cumbres Observatory Global Telescope Network Inc., \\6740 Cortona 
Drive Suite 102, Santa Barbara, CA 93117, USA \\
\vspace{2.5cm}
%\today 
}
\end{center}
\end{titlepage}
\newpage{\pagestyle{empty}\cleardoublepage}
\pagestyle{myheadings}
\markboth{}{}

\begin{center}
\LARGE{Temporal Variability of Stars and Stellar Systems}
\end{center}

\section{Introduction}
\protect\label{sec:Intro}

Although the Sun is our closest star by many orders of magnitude and despite
having sunspot records stretching back to ancient China, our knowledge of the
Sun's magnetic field is far from complete. Indeed, even now, after decades of
study, the most obvious manifestations of magnetic fields in the Sun (e.g.
sunspots, flares and the corona) are scarcely understood at all. These failures
in spite of intense effort suggest that to improve our grasp of magnetic fields
in stars and of astrophysical dynamos in general, we must broaden our base of
examples beyond the Sun;  we must study stars with a variety of ages, masses,
rotation rates, and other properties, so we can test models against as broad a
range of circumstances as possible.

Although optical interferometry continues to make great strides (e.g.
\citealt{monnier2007,zhao2008betalyr}), the tiny angular sizes of most stars
will make direct imaging of stellar surface features very difficult. This means
that we will have to rely on indirect methods to obtain information about the
surfaces of cool stars and their environment.  Over the next decade, this array
of techniques will be supplemented by rapidly maturing new capabilities such as
gyrochronology, asteroseismology and precision photometry from space, which will
transform our understanding of the temporal variability of stars and stellar
systems. In the next sections we will outline some of the key science questions
in this area along with the techniques that could be used to bring new insights
to these questions.

\section{Understanding the Rotation of Stars}
\protect\label{sec:RotEvol}

The surface rotation is an observable feature of stars that bears the imprint of
such fundamental processes as accretion, mass loss, time evolution of internal
flows and structure, and the action of magnetic fields.  Relating observed
phenomenology to underlying physics is difficult, however. For example, 
Solar-type stars with ages between that of the Hyades ($\sim625$\,Myr) and of
the Sun have observed rotation rates which are well-described by the Skumanich
$\omega \sim t^{-1/2}$ law (\citealt{soderblom1983}). But young clusters
($\sim50$--$100$\,Myr) show a  spread in rotation rates at a given mass,  with a
population of `ultrafast rotators' that are not seen in the older clusters 
(e.g. \citealt{stauffer1989alphaper}). Why do young stars show such a large
spread in rotation rate at each mass?  Models to address this question often
invoke saturation of the angular momentum loss above some critical angular
velocity $\omega_{sat}$. Yet there is evidence (e.g. \citealt{irwin2007ngc2516})
that even this is insufficient and that core-envelope rotational decoupling is
needed to explain the slowest rotators in open clusters. So what are the
important processes leading to the rotation distribution, and how do they work? 
Such questions have in the past been nearly unanswerable, for lack of relevant
observations and adequate supporting theory.  In the next decade, there is for
the first time an opportunity to address both of these failings.

Improved technology now allows major improvements in the fidelity, sample size,
and time coverage available to conventional kinds of observations. Thus, a
concerted effort using existing or planned wide-field imagers and multi-object
spectrographs would yield information of  unprecedented quality and 
homogeneity about rotation and magnetic fields on
large samples of cluster stars. Moreover, just as helioseismology has
revolutionized the understanding of the internal structure and rotation of the
Sun (e.g. \citealt{jcd2004solarrot}), asteroseismology is now poised to extend
this knowledge to a much greater range of stars. Asteroseismology-capable space
missions such as CoRoT and Kepler, as well as ground-based seismology networks
such as SONG (the Stellar Oscillations Network Group --
\citealt{grundahl2008song}) will yield meaningful estimates of stellar
convection zone depths (\citealt{monteiro2000, verner2006}), and also of the
depth dependence of the angular velocity in stars.

In addition to observational progress, great strides are expected in the
theoretical and computational resources that can be applied to the problem of
stellar rotation.  
Advances in 3-dimensional MHD models (e.g. \citealt{rempel2009}) combined
with insights about meridional  and other large-scale flows (e.g.
\citealt{jouve2008}) may provide a theoretical basis for interpreting the
anticipated new, highly detailed observational data.  This combination may for
the first time lead to a quantitative understanding of the rotation of stars.

\pagestyle{myheadings}
\markboth{Temporal Var. of Stars and Stellar Systems White Paper}{Temporal Var. of Stars and Stellar Systems White Paper}
\section{Determining the Ages of Stars}
\protect\label{sec:ages}

A star's age is one of its most fundamental parameters, yet it is the most
difficult to measure, particularly for objects in the field.  The internal
structure and composition of stars are expected to evolve on different paths
throughout the star's life, depending on the initial conditions.  But it is hard
to constrain evolutionary models relating observed stellar characteristics (such
as luminosity, radius, metallicity, rotation, chromospheric activity) with the
underlying physical processes (convection, mixing, radiative transport, magnetic
dynamo etc.) when the age of the subject cannot be established.  This flaw has
been sharply highlighted by recent exoplanet discoveries, in which the
corresponding theoretical work on the formation and evolution of planet-hosting
star systems \citep{melo2006,saffe2005} has been hampered by ignorance of
the host star's age.

In recent years, a new technique in stellar age-dating has become available, in
addition to improvements in the precision of traditional methods (e.g. measuring
the lithium depletion, chromospheric activity, asteroseismology and isochrone
fitting to HR diagrams \citealt{mamajek2007}).  \cite{barnes2007gyro}
established clear relationships describing stellar ages as a function of
rotation rates and star color.  Although currently the technique can be
applied only to solar-type (FGKM) stars over $\sim$200\,Myr whose rotation rates can
be measured, it can be used to date field stars.  

Currently, multiple age-dating techniques are being applied to clusters and
other targets for which the age can be corroborated (e.g. \citealt{meibom2008}). 
Confusingly, although the systems should be coeval, different methods
yield different ages. It is unclear whether this can be interpreted in terms
of a spread in the times of star formation.  In the near future, we anticipate
this work being extended to targets spanning wider ranges in age, metallicity
etc.  This will be complemented by advances in multidimensional hydrodynamical
simulations of stellar structure, which aim to improve our understanding of the
effects of physical processes such as rotation, convection, composition and
radiation transport on the accuracy of theoretical evolutionary tracks
\citep{mamajek2007}.

Over the next decade a number of keynote projects will provide unparalleled
survey data ideal for dating statistically significant samples of stars by
multiple methods.   The MOST (Microvariablity and Oscillations of Stars)
\citep{walker2003most} and CoRoT \citep{baglin2006seismo} missions are already
paving the way, providing high precision, continuously-sampled lightcurves of
thousands of stars.  The modulation of the lightcurves due to starspots
makes it easy to determine stellar rotation rates and hence apply
gyrochronology, in addition to asteroseismology techniques, providing a dual
constraint on age.  Kepler \citep{borucki2007kepler}, to be launched this year,
will monitor another extensive sample of $\sim$100,000 main sequence stars
(including 4 open clusters), many of which are expected to host planets. 
Furthermore, we look forward to the capabilities of the planned ground-based
1--2m-class telescope networks, such as SONG and LCOGT, to provide high cadence,
continuous, time series photometry and spectroscopy.  This will enable extensive
asteroseismology and stellar rotation studies of diverse stellar populations,
for which ages can now be derived.  

\section{Time Evolution of Stellar Activity}
\protect\label{sec:TimeEvol}

The Sun is variable in a wide variety of ways and on a large range of
timescales. Somewhat surprisingly given the length of records of sunspots, the
mean 11 year periodicity in sunspot number was identified only about a century and a
half ago \citep{schwabe1843,eddy1977}. Other periodicities have also been
detected and on top of this regular cyclic behavior are occasional absences of
any activity for prolonged lengths of time. However there is still considerable
controversy even for the Sun about how and where the field is generated and what
physical processes (differential rotation, meridional flow) control the magnetic
cycle. (e.g. \citealt{charbonneau2005solardynamo})

How can other stars help? Stars other than the Sun can be found at nearly any
age, mass and rotation period and provide a direct way of studying how the
dynamo processes depend on fundamental properties such as mass (and hence
convective zone depth) and rotation rate. It has been shown even for the `solar
twins' with long-term measurements \citep{hall2007}, that the dynamo can be very
sensitive to the basic parameters or rotation and mass and may depend on
additional parameters such as Li abundance. Clearly studying the dynamo
processes over a wide range of the potential parameter space is vital.

M dwarf stars are particularly interesting as they become fully convective (and
therefore lack an interface layer where dynamo processes concentrate in the Sun)
but nevertheless possess strong magnetic fields and activity (e.g.
\citealt{johnskrull1996, delfosse1998}). Part of this activity manifests itself
in the form of intense flares spanning the electromagnetic spectrum from the
X-ray to the radio. Since M dwarfs comprise about 70\% of stars in our Galaxy,
these flares will be a significant, even dominant, source of transients in the
time-domain surveys planned for the next decade, such as
PTF\footnote{\href{http://www.astro.caltech.edu/ptf/}{Palomar Transient Factory
(http://www.astro.caltech.edu/ptf/)}},
Pan-STARRS\footnote{\href{http://pan-starrs.ifa.hawaii.edu/public/}{http://pan-starrs.ifa.hawaii.edu/public/}}
and LSST\footnote{\href{http://www.lsst.org}{http://www.lsst.org}}. Clearly
determining the nature of the magnetic dynamo in low-mass stars will be crucial
for understanding the transient universe revealed by the synoptic surveys. But
beyond minimizing their nuisance to other science, the sources and mechanisms of
magnetic activity in M dwarfs is worth studying on its own terms, as a further
probe of hydromagnetic processes in stars. Previously this issue has been hard
to address, but new instruments and observations (\citealt{donati2008earlyM})
and improved theory and numerical simulation (\citealt{browning2008,ludwig2008})
will allow deeper understanding of these stars. 

\begin{figure}[ht]
\begin{centering}
\begin{tabular}{c}
\includegraphics[width=2.75in,height=4.5in,angle=270]{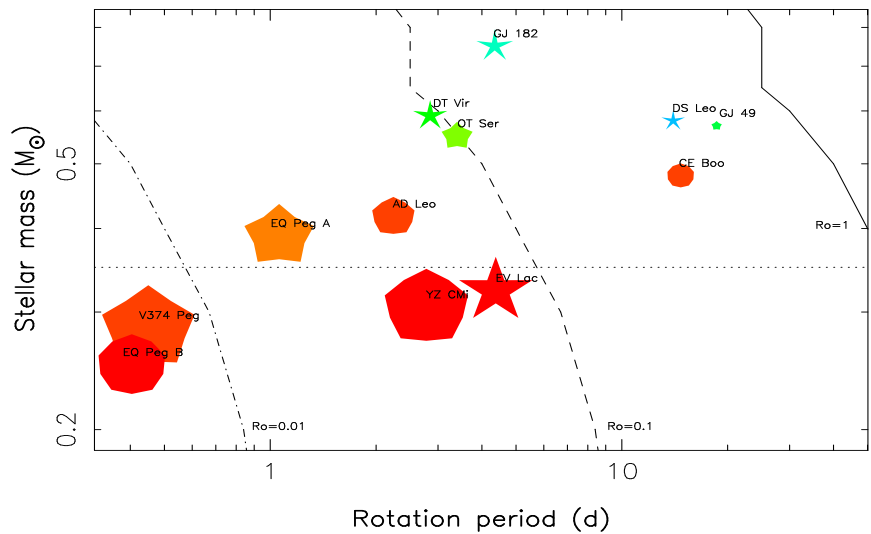}
\end{tabular}
\caption{Properties of large-scale magnetic properties of early and mid-M dwarfs
(from \citealt{donati2008earlyM}). Symbol size indicates field strength, symbol color indicates field configuration
(blue: purely toroidal, red: purely poloidal) and shape indicates degree of
axisymmetry (decagon: purely axisymmetric, stars: purely non-axisymmetric). The
theoretical fully-convective boundary is shown at $M_\star\simeq0.35 \mathrm{M}_\odot$.}
\protect\label{fig:dataflow}
\end{centering}
\end{figure}

Investments in new high resolution IR spectrographs and spectropolarimeters will
allow study of a large sample of M dwarfs with unprecedented accuracy and
sampling frequency.  In addition,
there will be a great increase in computing power over the next decade, allowing
numerical simulations to increase in spatial/wavelength resolution and to model
physical effects that are not currently included.  The resulting confrontations between
observation and theory may lead towards a coherent
picture of dynamo processes in stars.

\section{Connections to asteroseismology and to exoplanets}
\protect\label{sec:connect}

There are many other areas of astronomy and physics in general that overlap and
are influenced by the study of stellar magnetic fields. The
study of the Sun and its activity with facilities such as the
ATST\footnote{\href{http://atst.nso.edu/}{http://atst.nso.edu/}} and
SDO\footnote{\href{http://sdo.gsfc.nasa.gov/}{http://sdo.gsfc.nasa.gov/}}, along
with techniques like helioseismology, will allow magnetic fields in our
closest star to be studied in
great detail. 

The pace of progress in dynamo modeling accelerated after helioseismology
provided meaningful constraints on the Sun's interior structure and dynamics
\citep{brown1989sun}. Later observations, able to detect
helioseismic signatures of solar cycle effects, established that variations in
the mean strength of the solar magnetic field cause shifts of up to
($\sim$0.5~$\mu$Hz) in the frequencies of even the lowest-degree p-modes
(\citealt{libbrecht1990, salabert2004}). Space-based photometric asteroseismology missions,
such as MOST \citep{walker2003most}, CoRoT \citep{baglin2006seismo}, and Kepler
\citep{borucki2007kepler} will soon allow additional tests of dynamo models
using other solar-type stars (see \citealt{metcalfe2007},
\citealt{chaplin2007}). High precision time-series photometry from MOST has
already revealed latitudinal differential rotation in two solar-type stars
(\citealt{croll2006,walker2007}), and the long-term monitoring from future
satellite missions and from ground-based networks such as the Stellar Oscillations
Network Group (SONG) are expected to yield the precision necessary for
asteroseismic measurements of stellar convection zone depths
(\citealt{monteiro2000, verner2006}). By combining such observations with the
stellar magnetic activity cycles documented from long-term monitoring of the
Ca~{\sc ii} H and K lines, we can extend the calibration of dynamo models from
the solar case to a broad range of F, G, and K stars.
Adding interferometric measurements of stellar radii to this mix will
also be desirable, since this will strengthen constraints on fundamental
stellar properties such as mass (\citealt{creevey2007}).

Magnetic fields and temporal variability of stars
are also important because they affect conditions on any attendant planets.
The influence of the Sun on the Earth and its climate is of course
the subject of much work, and processes involving the solar wind and
UV flux have been critical in determining the evolution of the atmospheres
of Mars and Venus.
Stellar flares and other eruptive phenomena could have a similarly
large effect on the habitability of extrasolar planets,
especially those orbiting
highly variable M dwarfs, which are the focus of several
targeted planet search programs. 
Finally, magnetic activity is often the dominant noise source 
against which the radial velocity signal from extrasolar planets must be 
detected.
Better characterizing stellar activity and its radiative signatures
may be essential for isolating a numerous sample of planets circling
young, magnetically-active stars, and for confirming the many exoplanet
candidates expected from Kepler.

\section{Summary}
\protect\label{sec:conclusions}

Magnetic fields and dynamo processes play key roles in astrophysical
phenomena that range in scale from the dissipation length in flare
kernels up to the size of galaxies.
On Earth, a confluence of technical progress and broadening context has both
created opportunities and heightened motivation to mount a systematic
attack on the problem of stellar magnetic activity. 
For now, therefore, the most promising environment in which to study
these processes is on the intermediate scale, in stars.

If we could achieve a coherent picture of magnetic fields in stars, it
would illuminate many seemingly disparate phenomena.
Stellar activity cycles, stellar winds, angular momentum loss, and ages,
formation of young stars, irradiation and ablation of planetary atmospheres,
and high-energy flaring would then be treatable as closely-linked
and consistent processes.

Carrying out this line of research will require detailed study of the Sun,
using the most capable observing methods, the most innovative 
phenomenological models, and the highest-resolution numerical simulations.
It will also require a comprehensive extension of these methods to a wide
variety of other stars.
We know that different stars display a great range of differing
dynamo-related phenomena;
to capture the subtleties of the highly nonlinear magnetic interactions,
we will be forced to probe all of their behaviors.
Fortunately, the tools to do this are at hand.
Spaceborne photometry missions, ground-based robotic telescopes, and massive
expansion of computing power are the enabling technologies;
with adequate support,
international cooperation and imaginative scientific programs will provide
the needed organization.

% Uncomment the following (and matching \end) to produce two-column
% bibliography.
\begin{multicols}{2}
\bibliographystyle{abbrvnat}
\bibliography{WPrefs}
\end{multicols}

\end{document}